# Dimension Engineering of Narrow Bandgap Semiconductor InAs Nanostructures in Wafer-Scale

Dong Pan[1,5], Ji-Yin Wang[2], Wei Zhang[3], Lujun Zhu[4,7], Xiaojun Su[3], Shaoyun Huang[2], Manling Sui[4], Arkady Yartsev[3], H. Q. Xu[2] and Jianhua Zhao[1,5,6]★

[1]*State Key Laboratory of Superlattices and Microstructures, Institute of Semiconductors, Chinese Academy of Sciences, P.O. Box 912, Beijing 100083, China*

[2]*Beijing Key Laboratory of Quantum Devices, Key Laboratory for the Physics and Chemistry of Nanodevices and Department of Electronics, Peking University, Beijing 100871, China*

[3]*NanoLund and Division of Chemical Physics, Lund University, Box 124, 22100 Lund, Sweden*

[4]*Institute of Microstructure and Properties of Advanced Materials, Beijing University of Technology, Beijing 100124, China*

[5]*College of Materials Science and Opto-Electronic Technology, University of Chinese Academy of Sciences, Beijing, China*

[6]*CAS Center for Excellence in Topological Quantum Computation, University of Chinese Academy of Sciences, Beijing 100190, China*

[7]*College of Physics and Information Technology, Shaanxi Normal University, Xi'an 710062, China*

(Dated: November 12, 2018)

**Low-dimensional narrow bandgap III-V semiconductors are key building blocks for the next generation of high-performance nano-electronics, nano-photonics and quantum devices. To realize these various applications, it requires an efficient methodology that enables the material dimensional control during the synthesis process and the mass production of these materials with perfect crystallinity, reproducibility, low cost and outstanding optoelectronic properties. However, despite advances in one- and two-dimensional narrow bandgap III-V semiconductors synthesis, the progress towards reliable methods that can satisfy all of these requirements has been limited. Here, we demonstrate an approach that provides a precise control of the dimension of InAs from one-dimensional nanowires to wafer-scale free-standing two-dimensional nanosheets, which have a high degree of crystallinity and outstanding electrical and optical properties, using molecular-beam epitaxy by controlling catalyst alloy segregation. In our approach, two-dimensional InAs nanosheets can be obtained directly from one-dimensional InAs nanowires by silver-indium alloy segregation, which is much easier than all previously reported method, such as traditional buffering technique and select area epitaxial growth. Detailed transmission electron microscopy investigations provide a solid evidence that the catalyst alloy segregation is the origination of the InAs dimensional transformation from one-dimensional nanowires to two-dimensional nanosheets, and even to three-dimensional complex crosses. Using this method, we find that the wafer-scale free-standing InAs nanosheets can be grown on various substrates including Si, MgO, sapphire and GaAs *etc*. The InAs nanosheets grown at high temperature are pure phase single crystals and have a high electron mobility and a long time-resolved THz kinetics lifetime. Our work will open up a conceptually new and general technology route toward the effective controlling the dimension of the low-dimensional III-V semiconductors. It may also enable the low-cost fabrication of free-standing nanosheet-based devices on an industrial scale.**

Low-dimensional narrow bandgap III-V semiconductor InAs is an ideal material for applications in high-speed and low-power electronics, infrared optoelectronics and quantum-transport studies since it exhibits high electron mobility, narrow bandgap and strong spin-orbit coupling[1-3]. For example, the well-grown one-dimensional (1-D) InAs nanowires have been used to fabricate single-electron transistors, resonant tunneling diodes and ballistic transistors[4-6]. Recently, it has also been successfully used to detect the Majorana fermions in tunable semiconductor/superconductor hybrid devices[7-11]. In contrast to 1-D InAs nanowires, free-standing InAs nanosheets present another class of two-dimensional (2-D) material, where the band structure can be precisely tuned from bulk to 2-D by changing the thickness[12]. Its structural configuration also enables the direct contact of the active semiconductor layer with the gate stack, which is essential for enabling high-performance devices[13]. Very recently, 2-D InAs also shows great prospect in braiding or interferometric measurement on topological quantum computing[14]. All these applications require a high degree of InAs growth control on its morphology and especially dimension. Conventionally, a complex buffer layer engineering was used to grow 2-D InAs structures. Several reports have also been published on obtaining the 2-D InAs membranes and nanoleaves by modified epitaxial transfer scheme[12,15], select area epitaxial growth[16] and mirror-twin induced method[17]. However, the ability of these techniques to realize wafer-scale growth and well-controlled dimension modulation remains limited.

Here, we report a general method for growth of wafer-scale free-standing 2-D InAs nanosheets with controllable nanoscale dimensions, a high degree of crystallinity and outstanding electrical and optical properties using molecular-beam epitaxy (MBE) by catalyst alloy segregation. Our 2-D InAs nanosheets can be obtained directly from 1-D InAs nanowires by silver-indium alloy segregation, which is without any buffering technique and much easier than all previously reported method. The wafer-scale free-standing InAs nanosheets can be grown on various substrates including Si, MgO, sapphire and GaAs *etc*. The InAs nanosheets grown at high temperature are pure phase single crystals and have a high electron mobility and a long time-resolved THz kinetics lifetime. The growth method presented here not only provides a general and conceptually new approach for controlling the dimension of the low-dimensional III-V semiconductors, but also enables the low-cost fabrication of free-standing nanosheet-based devices on an industrial scale.

**Dimensional tunability of InAs from 1-D to 2-D**

We find that the dimension of InAs can be tuned directly from 1-D nanowires to 2-D nanosheets, and even to 3-D complex crosses and the indium flux is the only determined parameter for the dimensional tunability. Figure 1a-d is the schematic illustration of InAs with the morphology tuned from a 1-D nanowire (Fig. 1a) to a 2-D nanosheet (Fig. 1b), to a 1-D nanowire/2-D nanosheet cross (Fig. 1c), and to double 2-D nanosheets and one 1-D nanowire crosses (Fig. 1 d) by increasing the indium flux. Figure 1e-h shows corresponding typical transmission electron microscope (TEM) images for these different dimensional InAs nanostructures. Detailed morphology evolution process of InAs from 1-D to 2-D can be clearly observed from the samples grown at different indium fluxes (keeping the arsenic flux, growth temperature and growth time constants) on Si (111) substrates by MBE using silver as catalysts. As shown in Fig. 1i, for the sample grown with the indium flux of $2.0 \times 10^{-7}$ mbar, InAs nanowires are obtained on the substrate surface. Increasing the indium flux slightly up to $4.9 \times 10^{-7}$ mbar, sheet-like InAs nanostructures appear (Fig. 1j). Further increasing the indium flux to $7.4 \times 10^{-7}$ mbar, the resulting InAs nanostructures have obvious sheet morphology (Fig. 1k). By increasing the indium flux to $9.3 \times 10^{-7}$ mbar, high density and large size (length and width are both in micrometer scale) free-standing 2-D InAs nanosheets are obtained (Fig. 1l). At last, for the sample grown with a higher indium flux, 3-D complex InAs crosses appear. InAs planar epitaxial layers form on the substrates for the samples grown with very low arsenic/indium beam equivalent pressure (BEP) ratio (see Supporting Information Section S1). It is worth noting that the density of the 2-D InAs nanosheets increases remarkably with increasing the indium flux. Because the indium flux is the only parameter for the dimensional tunability, the dimensional transformation of InAs from 1-D to 2-D, and to 3-D should originate from the same crystal growth mechanism.

**Catalyst alloy segregation growth model**

As we observed in Fig. 1, the dimension of InAs can be tuned directly from 1-D nanowires to 2-D nanosheets, and to 3-D complex crosses just by tailoring the indium flux. To find out the reason for the dimensional evolution of InAs, spherical aberration-corrected (Cs-corrected) high angle annular dark field scanning transmission electron microscopy (HAADF-STEM) and energy-dispersive X-ray spectroscopy (EDS) mapping were performed on the InAs samples grown with different indium fluxes using an FEI Titan G2 microscope equipped with a super-X detector. For the sample grown with the indium flux of $2.0 \times 10^{-7}$ mbar, TEM results further confirm the sample are nanowires. As can be seen from Fig. 2a, InAs nanowires have homogeneous diameter, and spherical catalyst alloy particles at the end of the InAs nanowires can be clearly observed. High-resolution TEM and HAADF-STEM images of the spherical catalyst alloy particle indicate that the alloy particle is fully single-crystalline (Fig. 2b,c). EDS maps (Fig. 2e-h) and line scan (Fig. 2d) taken at the top region of the nanowire indicate that the remaining spherical catalyst particle is composed of silver and indium, and their elemental distributions are already a little bit inhomogeneous comparing with the samples grown with lower indium fluxes (see Supporting Information Fig. S2). As shown in Fig. 2i, increasing the indium flux slightly up to $4.9 \times 10^{-7}$ mbar, InAs has a sheet-like morphology, which is consistent with the scanning electron microscope (SEM) image in Fig. 1j. It is noteworthy that the spherical catalyst alloy particle is not single-crystal any more according to the high-resolution TEM and HAADF-STEM results in Fig. 2j,k. EDS maps (Fig. 2m-p) and line scan (Fig. 2l) taken at the top region of the InAs further indicate that the distribution of the silver atoms becomes inhomogeneous. Figure 2q is a typical TEM image of an InAs nanosheet grown with the indium flux of $7.4 \times 10^{-7}$ mbar. From the high-resolution TEM and HAADF-STEM results in Fig. 2r,s, we can see that the catalyst alloy particle is polycrystalline and the inhomogeneous distribution of the silver atoms in the catalyst alloy is more obvious. This phenomenon can be further confirmed by EDS maps (Fig. 2u-x) and the line scan (Fig. 2t). For the sample grown with the indium flux of $7.4 \times 10^{-7}$ mbar or higher, pure silver nanoparticles segregate from the catalyst alloy of InAs, and they can be clearly observed and detected from the STEM images and EDS point analysis (see Supporting Information Section S2). For the samples grown with a lager indium flux, thin indium shells form on the surface of catalyst alloy particles and InAs nanosheets (Fig. 2h,p,x), which confirm that the indium-rich condition plays an important role in the dimensional transformation of InAs.

Because the indium-rich condition is the only requirement for the catalyst alloy segregation as well as the dimensional evolution of InAs from 1-D to 2-D, the catalyst alloy segregation process appeared in the samples grown with different indium fluxes (Fig. 2a-x and Figs. S2 and S3) should be observed for a sample grown under the indium-rich condition at different growth stages. To further figure out the catalyst alloy segregation process for the dimensional evolution of the InAs from 1-D to 2-D, a set of InAs samples have been grown under the same indium-rich growth condition with different growth time. Figure 3a-f is side-view SEM images of InAs grown on Si (111) substrates with growth time of 2.5 min, 5 min, 10 min, 15 min, 40 min and 80 min, respectively. As shown in Fig. 3a, the sample grown for 2.5 min, the InAs is the 1-D nanowires and their lengths are from several tens to 100 nm. Extending the growth time to 5 min, the length of InAs increases to about 500 nm, and the morphology of InAs is still the 1-D nanowire with a homogenous diameter (Fig. 3b), but the droplets on the top of nanowires are very unstable (see Supporting Information Section S3). When the growth time is increased to 10 min, it can be found that the length of InAs nanowires further increases. However, the nanowires have an obviously lateral growth trend (Fig. 3c). Further increasing the growth time to 15 min, sheet-like InAs can be observed (Fig. 3d). For even longer growth time (40 min), larger-scale high density free-standing 2-D nanosheets are grown on the substrate (Fig. 3e and Supporting Information Section S4). When the growth time is extending to 80 min, 2-D nanosheets with a larger-scale, super-high density, free-standing and lager size (length and width can be up to 10 μm and 2 μm, respectively) are obtained, as observed in Fig. 3f. The density of the 2-D nanosheets on the substrate surface increases with increasing sample growth time, which indicates that the indium-rich growth condition is very suitable for the 2-D nanosheet growth as we observed in Fig. 1i-l. For this series InAs samples, detailed TEM, HAADF-STEM and EDS results confirm that the catalyst alloy on top of InAs has undergone the same segregation process (see Supporting Information Section S5) as that in the InAs samples grown with the different indium fluxes in Fig. 1.

According to above results, we believe that the dimensional evolution of InAs from 1-D to 2-D is oriented from the catalyst alloy segregation under an indium-rich growth condition. The schematic demonstration of the catalyst alloy segregation process under the indium-rich condition is given in Fig. 3g-o. First, if InAs grown under an indium-poor growth condition or located at the initial stage of the growth as shown in Fig. 3a (the value of indium flux is smaller than that can produce an unstable droplet), the morphology is 1-D nanowire with a single crystal spherical silver-indium alloy droplet on its top (Fig. 3g). At this growth condition, the interface between nanowire and catalyst alloy nanoparticle is sharp as that in the III-V semiconductor nanowires seeded by other metals[18,19] due to the isotropical growth of the nanowires. Then, when the growth atmosphere of InAs is gradually becoming indium-rich by increasing the indium flux (or increasing the growth time in a large indium flux), the silver-indium alloy droplet will become unstable and the distribution of the silver atoms becomes inhomogeneous (Fig. 3h). As soon as the indium-rich growth condition reaches, the silver-indium alloy droplet starts segregation and the morphology of InAs evolves from 1-D nanowire to 2-D nanosheet gradually due to the anisotropical growth caused by catalyst alloy segregation (Fig. 3i). As the growth and segregation proceeds, the silver-indium alloy droplet finishes segregation and the segregated silver droplet can migrate to the substrate surface as shown in Fig. 3j (If silver droplet migrates onto the surface of nanosheet, the sample will transform from 1-D nanowires to 3-D complex crosses, see Supporting Information Section S6). The interface between 2-D nanosheet and the catalyst alloy nanoparticle is not sharp any more owing to the above anisotropical growth. After that, the silver-indium alloy droplet on the 2-D nanosheet becomes unstable, and the segregated silver droplet seeds a new InAs nanowire growth (Fig. 3k and Fig. S10). As the growth continues, the silver-indium alloy droplet on the 2-D nanosheet starts segregation again, and the silver-indium alloy droplet on the new 1-D InAs nanowire also becomes unstable (Fig. 3l). Next, as can be seen from Fig. 3m, the silver-indium alloy droplet on the 2-D nanosheet finishes segregation and the segregated silver droplet can also migrate to the substrate surface. The silver-indium alloy droplet on the new 1-D InAs nanowire also starts segregation and the morphology of the new InAs evolves from 1-D nanowire to 2-D nanosheet gradually. After that, silver-indium alloy droplets continuously undergo the same segregation process shown in Fig. 3g-m and high density 2-D InAs nanosheets (Fig. 3n,o) form on the substrate. The catalyst alloy segregation takes place in all the samples grown with different

V/III BEP ratios as long as the growth condition is indium-rich. For samples grown under the same large indium flux with different growth time, the growth condition will also become indium-rich more and more with the increasing of the growth time. The more the indium-rich condition, the faster the segregation process. That is why we have observed the dimension of InAs can be tuned directly from 1-D to 2-D, and the density of the 2-D InAs nanosheets both increase remarkable with increasing indium flux (Fig. 1) and growth time (Fig. 3). With continuing the segregation process, the volume of the droplets deduces gradually. Thus, very thin 2-D InAs nanosheets can be obtained in our samples.

**Generality and scope of catalyst alloy segregation crystal growth**

We find that the tuning the dimension of the low-dimensional III-V semiconductors by the catalyst alloy segregation is a general method. For example, we can realize the dimensional tunability of InAs from 1-D to 2-D using 'catalyst alloy segregation' with a two-step growth procedure. In the so-called two-step growth process, InAs nanowires are firstly grown on the substrate with silver as catalysts (Fig. 4a,b), and then tuning the growth conditions to 'indium-rich' (after InAs nanowire growth, increasing the indium flux to a large value and keeping other parameters constant) to ensure that the catalyst alloy segregation can happen. As shown in Fig. 4c-i, once the catalyst alloy segregation happens, the morphology of InAs will evolve from 1-D nanowire to 2-D nanosheet with the same process described in Fig. 3g-o. Figure 4m-o is the SEM images of the 2-D InAs nanosheets grown on the 1-D nanowires with the two-step growth procedure. All the InAs nanowires are firstly grown with the V/III BEP ratio of 39.3, growth time of 40 min and growth temperature of 505 °C (see Supporting Information Section S7). In the second step, samples are then grown with different indium fluxes and growth temperatures, respectively. We can see from Fig. 4m,n that 2-D InAs nanosheets can be obtained with the V/III BEP ratios of 12.0 (Fig. 4j) and 7.0 (Fig. 4k). The density and size of the former are obvious larger than the later owing to the larger indium flux used, which is consistent with the results obtained directly on the Si substrates in Fig. 1. As shown in Fig. 4o, 2-D InAs nanosheets can also be observed on the substrate for the sample grown with a higher temperature (525°C, Fig. 4l). To exclude the possibility of the 2-D InAs nanosheets are obtained from lateral VS growth owing to the possible axial growth stopped by the growth condition variation, the second-step growth has been carried out on the InAs nanowires with short length (e.g. InAs nanowires grown with 10 min and 15 min), and it is found that the high density and large size 2-D InAs nanosheets can still be obtained (see Supporting Information Section S7).

Another very important characteristic for the 'catalyst segregation crystal growth' is that wafer-scale free-standing 2-D InAs nanosheets can be realized by this method on various substrates with different size including Si, MgO, sapphire and GaAs *etc.* Figure 5a-d shows the typical top view substrate images taken before and after sample grown on the 2 inch p-type Si (111), 2 cm × 2 cm MgO (100) and 2 inch polished and 3 inch unpolished sapphire (0001) substrates, respectively. As shown in Fig. 5a, Si (111) substrate has a clean surface (like a mirror) before the 2-D InAs nanosheet growth. In contrast, the substrate has a very homogenous black surface after sample growth. MgO (Fig. 5b) and sapphire (Fig. 5c,d) substrates are transparent before the sample growth, while they are all homogenous black in color after sample growth. Interestingly, 2-D InAs nanosheets are not sensitive to the substrates surface, as can be seen from Fig. 5c and Fig. 5d, samples grown on the polished and unpolished sapphire (0001) substrates both have homogenous color. Detailed high amplified SEM images indicate that 2-D InAs nanosheets have very homogeneous density and distribution in the different area of all these substrates (Supporting Information Section S8). Here, the dimensional tunability of wafer-scale InAs from 1-D to 2-D at various substrates with different size can be expected substantially to reduce the cost of producing high-quality low-dimensional InAs and may enable the low-cost fabrication of InAs-based devices on an industrial scale.

**Thickness-modulated and phase purity of 2-D InAs nanosheets**

The thickness of the 2-D InAs nanosheets is determined by the diameter of the 1-D InAs nanowires. For the 2-D InAs nanosheets obtained directly on the substrates as mentioned above, the thickness of the 2-D InAs nanosheets can be tuned by the silver catalyst diameter. Very thin 2-D InAs nanosheets with thickness less than 10 nm can be evolved from the ultrathin 1-D InAs nanowires. In contrast, the thick 2-D InAs nanosheets with thickness large than 100 nm can be obained from the large diameter 1-D InAs nanowires. For the 2-D InAs nanosheets grown with the two-step growth procedure, their thickness can be tuned with the same way (see Supporting Information Section S9).

As to the crystal structure of the 2-D InAs nanosheets, the nanosheets are composed of a mixture of wurtzite (WZ) and zinc-blende (ZB) segments (Fig. 2j,r,k,s) as reported in most III-V 1-D nanowires and 2-D nanomembranes grown with a bottom- up manner[20,16]. These effects might have an impact on the nanosheets' optical and electrical properties and may pose problems in future nano-electronic devices due to electron scattering at stacking faults or twin planes[21,22]. Careful control of the structure of III-V nano-materials will improve the success of future devices, including heterostructure devices. In our work, 2-D InAs nanosheets can be well grown with a large substrate temperature range (see Supporting Information Section S10). Luckily, we find that phase pure 2-D InAs nanosheets can be achieved by only varying the growth temperature. Figure 6 are the typical TEM images of InAs nanosheets grown at different temperatures. For nanosheets grown at 475 °C (Fig. 6a), the structure is a mixture of WZ and ZB phases. As shown in Fig. 6b-d, stacking faults and twin defects can be clearly observed from the HRTEM images taken from top and side regions in the nanosheet. In Fig. 6e-h, it is observed that the density of single stacking faults is obviously reduced for the nanosheet grown at 475 °C. Further increasing the growth temperature to 525 °C (Fig. 6i), we can see that the density of stacking faults is strongly reduced. The number of the stacking faults is countable using the dark-field TEM (Fig. 6j) and HRTEM images (Fig. 6k,l). At high growth temperatures (545 °C), selected area electron diffraction (SAED, Fig. 6q-t) patterns and HRTEM images (Fig. 6o,p) confirm that nanosheets have high crystal-quality with only one or nearly no stacking faults and other defects along their entire length (see Supporting Information Section S11). Our results are accordance with the previous report on a high growth temperature coupled with a low V/III BEP ratio (indium-rich condition) produced pure WZ nanowires free of stacking faults[23]. We find that the V/III BEP ratio (indium flux in this work) is not very sensitive to the crystal-quality of the 2-D InAs nanosheets. As shown in Fig. 2, the structures are all composed with WZ and ZB mixture phases with the indium fluxes increasing from $4.9\times10^{-7}$ mbar to $9.3\times10^{-7}$ mbar. To our surprising, if we further increasing the indium flux to $1.1\times10^{-6}$ mbar (V/III=5.4), the crystal-quality can be much improved even for the sample grown at a low temperature (505 °C, see Supporting Information Section 12). Phase purity achieved without sacrificing important specifications of size and dopant level opens new possibilities for engineering nanosheet devices, without restrictions on nanosheet size or doping.

**Electrical and optical properties of 2-D InAs nanosheets**

In order to access the electronic properties of the grown InAs nanosheets, we have performed transport measurements on nanosheet devices. It is noted that in this paper we have measured three kinds of devices labeled as A, B and C, in which the used InAs nanosheets are grown in three different conditions. For A devices, InAs nanosheets are grown at 455 °C with a V/III BEP ratio of 6.3. For B devices, the growth temperature is increased to 545 °C and the V/III BEP ratio is fixed at 6.3. For C devices, InAs nanosheets are grown at 505 °C with a V/III BEP ratio of 5.4. Figure 7a shows the atomic force microscopy (AFM) image of device B-01. In the device, the source and drain electrodes are fabricated by depositing 5-nm titanium and 90-nm gold. Details about the fabrication processes and device geometries are shown in Supporting Information Section 13. As shown in Fig. 7b, a bias voltage $V_{sd}$ is applied on the source with drain grounded and the carrier density in InAs nanosheet is tuned by back-gate voltage $V_g$. Figure 7c shows the output characteristic curves of device B-01 at different back-gate voltages at $T$=2 K. The linear output curves indicate that the device possesses ohmic contact between electrodes and nanosheet even at a low temperature. Figure 7d depicts transfer characteristic curves of device B-01 with $V_{sd}$=10 mV at different temperatures. The device behaves as a typical n-type field-effect transistor. We find that with decreasing temperature the transconductance becomes steeper implying that electron mobility is improved. By fitting the transfer characteristic curves (see details in Supporting Information Section 13), we can extract field-effect mobility of the InAs nanosheet devices as shown in Fig. 7e. By comparing A devices with B devices, we find that B devices possess much higher electron mobility globally at all accessible temperatures, indicating that the InAs nanosheets of B devices possess much higher crystalline quality than that of A devices. The results are consistent with TEM and SAED results of Fig. 6 that increasing growth temperature can improve the crystalline quality of InAs nanosheets. In Fig. 7e, we observe that at room temperature the electron mobility of B devices is about ~2000 cm$^2$/V•s, much larger than other un-capsuled two-dimensional material such as layered MoS$_2$[24] and black phosphorus[25]. At $T$=10 K, the electron mobility of B devices can reach to ~7000 cm$^2$/V•s, almost as high as InSb nanosheets[26]. Besides, in Fig. 7e we find that the mobility of C devices is nearly close even if not as high as that of B devices, implying that the crystalline quality of InAs nanosheets grown at low temperature can be improved by further reducing the V/III BEP ratio, which is consistent the TEM results in Supporting Information Section 12. Such high-quality 2-D nanosheets can work as a building block for constructing scalable high-performance nano-electronic and quantum devices in the future.

Time-resolved THz spectroscopy (TRTS) is selectively sensitive to photoconductivity of the samples, which is propotional to the product of mobility and concentration of photo-generated charges. Consequently, TRTS kinetics can be used to analyse carrier recombination processes as well as mobility dynamics. In planar InAs, electron mobility is significantly higher than that of holes[27,28], thus we presume that photoconductivity is mainly arising from photogenerated electrons. In InAs nanosheets, the photoconductivity ($\Delta\sigma$) dynamics can be described by the following rate equation[29]:

$$\frac{d(\Delta\sigma)}{dt} = e\mu_e(t) \times (-AN(t)^3 - \gamma N(t)^2 - \frac{N(t)}{\tau_{e-trap}}),$$

where $\mu_e(t)$ is the electron mobility, $N$ is the concentration of electrons, $A$ and $\gamma$ are the rates of the Auger and bi-molecular recombination, respectively and $1/\tau_{e-trap}$ is the electron trapping rate. TRTS kinetics of InAs nanosheets with varied growth temperatures were measured by the setup described in Fig. 8a. As shown in the Fig. 8b, the $\Delta\sigma$ decays of 1-D nanowires and 2-D nanosheets can be fitted by single exponential decay functions, suggesting that TRTS kinetics are dominated by monomolecular recombination processes such as electron trapping[29]. The fitted lifetimes for 1-D nanowires is 49.7±4.8 ps, while for 2-D nanosheets are 258.9±7.8, 310.2±13.6 and 354.4±23.8 ps for samples grown at 475 °C (blue circle), 505 °C (green circle) and 525 °C (red circle), respectively. Apparently, the $\Delta\sigma$ lifetime of 2-D nanosheets is much longer than that of 1-D nanowires and gradually increases with the growth temperature. In high crystallinity semiconductors, electron mobility does not change with time, thus the variation of the lifetime is caused by different charge trapping. Therefore, slower decay of photoconductivity can be attributed to the reduced trap density in the nanosheets with increasing growth temperature. In 2-D InAs nanosheets, both the surface and bulk traps can contribute to the electron trapping processes. To understand the nature of traps, we examined SEM and TEM of the samples. From SEM, we find that the overall geometric structure of nanosheets grown at 475 °C, 505 °C and 525 °C are similar. Accordingly, we would expect similar surface trap density of samples with varied growth temperatures, and TRTS decays are most likely dominated by electron bulk traps, which decrease with increasing sample growth temperature. These results agree with TEM measurements above.

**Conclusions**

In summary, we demonstrate a new and general strategy to realizing the wafer-scale dimensional tunability of low-dimensional III-V semiconductors. Wafer-scale free-standing 2-D InAs nanosheets with controlled dimensions, a high degree of crystallinity and outstanding electrical and optical properties have been obtained using MBE by the catalyst alloy segregation. The catalyst alloy segregation can be used to tune the InAs from 1-D nanowires to 2-D nanosheets, and even to 3-D complex crosses. The wafer-scale free-standing 2-D InAs nanosheets can be grown on various substrates including Si, MgO, sapphire and GaAs *etc*. The phase purity of 2-D InAs nanosheets can be achieved for the sample grown at high temperature. These pure phase single crystal InAs nanosheets have a high electron mobility and a long time-resolved THz kinetics lifetime. Although in this work we have respectively focused on InAs and silver as the semiconductor and catalyst, other compound semiconductors and metal catalysts could be explored in the future. The proposed nanosheet growth mechanism could be general to many materials. The reported wafer-scale and potentially high throughput method can be expected substantially to reduce the cost of producing high-quality nanosheets and may enable the low-cost fabrication of nanosheet-based devices on an industrial scale. At the moment our results cannot be fully explained by the existing literature, and we expect these results to inspire the development of new theoretical models.

**Methods**

All InAs nanostructures were grown in a solid source MBE system (VG 80) using silver as seed particles. For catalyst deposition, we used a silver effusion cell attached to the III-V growth chamber. This configuration enabled us to deposit the catalyst on a chemically clean surface and at the same time to control the substrate temperature and monitor the deposition process with reflection high-energy electron diffraction. The conventional one step catalyst annealing process was used to generate silver nanoparticles as reported in our previous work[30]. Commercial p-type Si (111), MgO (100), polished and unpolished sapphire (0001) and GaAs (100) were used as the substrates. Before loading the substrates into the MBE chamber, they were cleaned and treated using different methods to remove the surface contamination or native oxide. The detailed substrate treating processes and sample growth parameters were given in the Supplementary Information. Samples were grown at V/III BEP ratios from 3.0 to 29.5 (by increasing the indium flux while keeping the arsenic ($As_4$) flux constant) with growth time of 40 min and growth temperature of 505 °C in the variant indium flux series (see Supporting Information Section S1 and S2). Samples were grown at temperatures from 420 °C to 545 °C with growth time of 40 min and V/III BEP ratio of 6.3 in the variant growth temperature series (see Supporting Information Section S10). Samples were grown at time from 2.5 to 80 min with a V/III BEP ratio of 6.3 and growth temperature of 505 °C in the variant growth time series (see Supporting Information Section S4).

To realize the 2-D InAs nanosheets directly epitaxially grown along 1-D InAs nanowire frameworks, the InAs nanowires were firstly grown on the p-type Si (111) substrates with V/III BEP ratio of 39.3, growth temperature of 505 °C and growth time of 40 min. The InAs nanowire growth was terminated by closing the indium source shutter and keeping the arsenic source shutter open. Following that, the V/III BEP ratio and growth temperature were varied. Finally, the 2-D InAs nanosheets growth was initiated by opening the indium source shutter (see Supporting Information Section S7). The growth conditions of the 2-D InAs nanosheets grown on MgO (100), sapphire (0001) and GaAs (100) substrates were same to that on the Si (111) substrates (see

Supporting Information Section S8). All the InAs nanostructure growth was terminated by switching off the indium supply while maintaining the arsenic supply until the substrate was cooled down below 300 $^0$C within a few minutes.

The morphologies of the samples were observed by SEM with a Nova NanoSEM 650, operated at 20 kV. Structural characterization was performed using TEM and samples were removed from the growth substrate via sonication in ethanol and then drop-cast onto a holey carbon film supported by a copper grid. High-resolution TEM, HAADF-STEM images and EDS spectra (including the EDS elemental mapping and line scans shown in Fig. 2 in main text and Supporting Information Section S5) were taken with an FEI Titan G2 microscope equipped with a super-X detector, operated at 300 kV. Other high-resolution TEM images were collected using a JEOL2100, operated at 200 kV.

The detailed device fabrications and device parameters of the 2-D InAs nanosheets were provided in Supporting Information Section S13. The time resolved THz measurements were conducted by a setup described in Refs 31 and 32. Laser pulses (796 nm, 80 fs pulse length, 1 kHz repetition rate) were generated by a regenerative amplifier (Spitfire Pro XP, Spectra Physics) seeded by a femtosecond oscillator (MaiTai, Spectra Physics). The laser beam was split into three. The first beam (400 µJ/pulse) was used to generate THz radiation by optical rectification in a MgO: LiNbO$_3$ crystal. The second beam was used for electro-optical sampling of the THz pulses in a (110) ZnTe crystal. The third beam, which was used for excitation was converted to the second harmonic (398 nm) in a BBO crystal. To avoid absorption of the THz radiation by water vapor, the setup was purged with dry nitrogen. To conduct TRTS measurement, nanowire and nanosheet samples were embedded in the colorless First Contact polymer (CFCP) film (Photonic Cleaning Technologies LLC) and peeled off from native substrates. CFCP film is an optically transparent, inert, plastic film, and no photoconductivity response of CFCP was observed in the reference TRTS measurement.

## Author contributions

D.P. and J.H.Z. designed the experiments. D.P. grew all the samples. D.P., L.J.Z. and M.L.S analysed the structures and compositions using TEM and EDS. J.Y.W, S.Y.H. and H.Q.X. fabricated the devices, performed the electrical measurements and analysed the electrical data. W.Z., X.J.S and A.Y. performed the optical measurements and analysed the optical data. D.P. and J.H.Z. wrote the manuscript. All authors discussed the results and commented on the manuscript.

★To whom correspondence should be addressed.
E-mail: jhzhao@red.semi.ac.cn

Notes
The authors declare that they have no competing financial interests.

## ACKNOWLEDGMENTS
This work was supported by the MOST of China (Grant No. 2015CB921503), the National Natural Science Foundation of China (Grant Nos. 61504133 and 61334006) and the Strategic Priority Research Program of Chinese Academy of Sciences (Grant No. XDB28000000). D.P. also acknowledges the support from Youth Innovation Promotion Association, Chinese Academy of Sciences (No. 2017156).

## REFERENCES

1. del Alamo, J. A. Nanometre-scale electronics with III-V compound semiconductors. *Nature* **479,** 317 (2011).
2. Tomioka, K.; Yoshimura, M.; Fukui, T. A III-V nanowire channel on silicon for high-performance vertical transistors. *Nature* **488,** 189 (2012).
3. Pettersson, H., Trägårdh, J., Persson, A. I., Landin, L., Hessman, D. and Samuelson, L. Infrared photodetectors in heterostructure nanowires. *Nano Lett.* **6,** 229 (2006).
4. Thelander, C., Mårtensson, T., Björk, M. T., Ohlsson, B. J., Larsson, M. W., Wallenberg, L. R., Samuelson, L. Single-electron transistors in heterostructure nanowires. *Appl. Phys. Lett.*, **83,** 2052 (2003).
5. Björk, M. T., Ohlsson, B. J., Thelander, C., Persson, A. I., Deppert, K., Wallenberg, L. R., Samuelson, L. Nanowire resonant tunneling diodes. *Appl. Phys. Lett.* **81,** 4458 (2002).
6. Chuang, S., Gao, Q., Kapadia, R., Ford, A. C., Guo, J., Javey, A. Ballistic InAs nanowire transistors. *Nano Lett.* **13,** 555 (2013).
7. Das, A. et al. Zero-bias peaks and splitting in an Al-InAs nanowire topological superconductor as a signature of Majorana fermions. *Nature Phys.* **8,** 887 (2012).
8. Krogstrup, P., Ziino, N. L. B., Chang, W., Albrecht, S. M., Madsen, M. H., Johnson, E., Nygård, J., Marcus, C.M., Jespersen, T. S. Epitaxy of semiconductor-superconductor nanowires. *Nature*

*Mater.* **14,** 400 (2015).
9. Albrecht, S. M., Higginbotham, A. P., Madsen, M., Kuemmeth, F., Jespersen, T. S., Nygård, J., Krogstrup, P., Marcus, C. M. Exponential Protection of Zero Modes in Majorana Islands. *Nature* **531,** 206 (2016).
10. Deng, M. T., Vaitiekėnas, S., Hansen, E. B., Danon, J., Leijnse, M., Flensberg, K., Nygård, J., Krogstrup, P., Marcus, C. M. Majorana bound state in a coupled quantum-dot hybrid-nanowire system. *Science* **354,** 1557 (2016).
11. Kang, J. H., Grivnin, A., Bor, E., Reiner, J., Avraham, N., Ronen, Y., Cohen, Y., Kacman, P., Shtrikman, H. and Beidenkopf, H. Robust epitaxial Al coating of reclined InAs nanowires. *Nano Lett.* **17,** 7520 (2017).
12. Takei, K., Fang, H., Kumar, S. B., Kapadia, R., Gao, Q., Madsen, M., Kim, H. S., Liu, C.-H., Chueh, Y.-L. Plis, E., Krishna, S., Bechtel, H. A., Guo, J., Javey, A. Quantum confinement effects in nanoscale-thickness InAs membranes. *Nano Lett.* **11,** 5008 (2011).
13. Rogers, J. A., Lagally, M. G., Nuzzo, R. G. Synthesis, assembly and applications of semiconductor nanomembranes. *Nature* **477,** 45 (2011).
14. Suominen, H. J., Kjaergaard, M., Hamilton, A. R., Shabani, J., Palmstrøm, C. J., Marcus, C. M., and Nichele, F. Zero-energy modes from coalescing nndreev states in a two-dimensional semiconductor-superconductor hybrid platform. *Phys. Rev. Lett.* **119,** 176805 (2017).
15. Ko, H., Takei, K., Kapadia, R., Chuang, S., Fang, H., Leu, P. W., Ganapathi, K., Plis, E., Kim, H. S., Chen, S.-Y., Madsen, M., Ford, A. C., Chueh, Y.-L., Krishna, S., Salahuddin, S., Javey, A. Ultrathin compound semiconductor on insulator layers for high-performance nanoscale transistors. *Nature* **468,** 286 (2010).
16. Conesa-Boj, S., Russo-Averchi, E., Dalmau-Mallorqui, A., Trevino, J., Pecora, E. F., Forestiere, C., Handin, A., Ek, M., Zweifel, L., Wallenberg, L. R., Rüffer, D., M. Heiss, Troadec, D., Negro, L. D., Caroff, P. and i Morral, A. F. Vertical "III-V" V-shaped nanomembranes epitaxially grown on a patterned Si [001] substrate and their enhanced light scattering. *ACS Nano* **6,** 10982 (2012).
17. Soo, M. T., Zheng, K., Gao, Q., Tan, H. H., Jagadish, C., and Zou, J. Mirror-twin induced bicrystalline InAs nanoleaves. *Nano Res.* **9,** 766 (2016).
18. Caroff, P., Dick, K. A., Johansson, J., Messing, M. E., Deppert, K., Samuelson, L. Controlled polytypic and twin-plane superlattices in III-V nanowires. *Nat. Nanotech.* **4,** 50 (2009).
19. Dick, K. A., Caroff, P. Metal-seeded growth of III-V semiconductor nanowires: towards gold-free synthesis. *Nanoscale* **6**, 3006 (2014).
20. Hjort, M., Lehmann, S., Knutsson, J., Zakharov, A., Du, Y. A., Sakong, S., Timm, R., Nylund, G., Lundgren, E., Kratzer, P., Dick, K. A., and Mikkelsen, A. Electronic and structural differences between wurtzite and zinc blende InAs nanowire surfaces: experiment and theory. *ACS Nano* **8,** 12346 (2014).
21. Stiles, M. D., Hamann, D. R. Electron transmission through silicon stacking faults. *Phys. Rev. B* **41,** 5280 (1990).
22. Stiles, M. D., Hamann, D. R. Ballistic electron transmission through interfaces. *Phys. Rev. B* **38,** 2021 (1988).
23. Joyce, H. J., Wong-Leung, J., Gao, Q., Tan, H. H., and Jagadish, C. Phase perfection in zinc blende and wurtzite III-V nanowires using basic growth parameters. *Nano Lett.* **10,** 908 (2010).
24. Wang, Q. H., Kalantar-Zadeh, K., Kis, A., Coleman, J. N., Strano, M. S. Electronics and optoelectronics of two-dimensional transition metal dichalcogenides. *Nat. Nanotech.* **7,** 699 (2012).
25. Li, L. *et al.* Black phosphorus field-effect transistors. *Nat. Nanotech.* **9**, 372 (2014).
26. Pan, D., Fan, D. X., Kang, N., Zhi, J. H., Yu, X. Z., Xu, H. Q., Zhao, J. H. Free-standing two-dimensional single-crystalline InSb nanosheets. *Nano Lett.* **16,** 834 (2016).
27. Karataev, V. V., Mil'vidskiî, M. G., Rytova, N. S., Fistul', V. I. Compensation in n-type InAs. *Sov. Phys. Semicond.* **11,** 1009 (1977).
28. Nainani, A., Bennett, B. R., Brad Boos, J., Ancona, M. G., Saraswat, K. C. Enhancing hole mobility in III-V semiconductors. *J. Appl. Phys.* **111,** 103706 (2012).
29. Boland, J. L., Amaduzzi, F., Sterzl, S., Potts, H., Herz, L. M., i Morral, A. F., Johnston, M. B. High electron mobility and insights into temperature-dependent scattering mechanisms in InAsSb nanowires. *Nano Lett.* **18,** 3703 (2018).
30. Pan, D., Fu, M. Q., Yu, X. Z., Wang, X. L., Zhu, L. J., Nie, S. H., Wang, S. L., Chen, Q., Xiong, P., von Molnár, S., Zhao, J. H. Controlled synthesis of phase-pure InAs nanowires on Si (111) by diminishing the diameter to 10 nm. *Nano Lett.* **14,** 1214 (2014).
31. Ponseca, Jr., C. S., Němec, H., Wallentin, J., Anttu, N., Beech, J. P., Iqbal, A., Borgström, M., Pistol, M.-E., Samuelson, L., Yartsev, A. Bulk-like transverse electron mobility in an array of heavily n-doped InP nanowires probed by terahertz spectroscopy. *Phys. Rev. B* **90,** 085405 (2014).
32. Zhang, W.; Zeng, X.; Su, X.; Zou, X.; Mante, P.-A.; Borgström, M. T.; Yartsev, A. Carrier recombination processes in gallium indium phosphide nanowires. *Nano Lett.* **17,** 4248 (2017).

**Figures and captions**

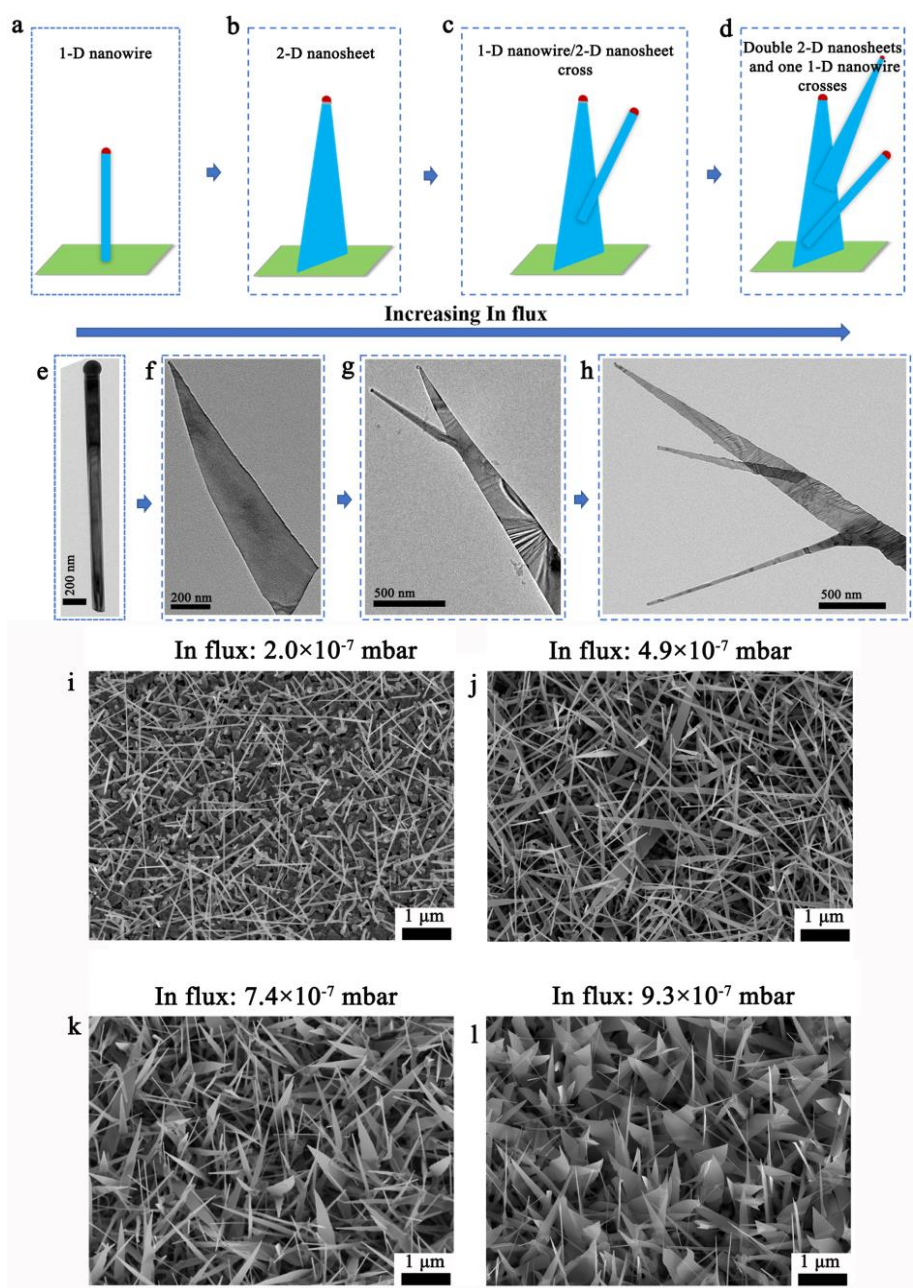

**Figure 1 | Dimensional tunability of InAs from 1-D nanowires to 2-D nanosheets and to 3-D complex crosses by increasing the indium flux. a-d**, Schematic illustration of InAs with the morphology tuned from 1-D nanowire (a) to 2-D nanosheet (b), to 1-D nanowire/2-D nanosheet cross (c), and to double 2-D nanosheets and one 1-D nanowire crosses (d) by increasing the indium flux. **e-h**, TEM images of a typical 1-D nanowire, 2-D nanosheet, 1-D nanowire/2-D nanosheet cross and double 2-D nanosheets and one 1-D nanowire crosses, respectively. **i-l**, 25° tilted SEM images of InAs nanostructures grown at different indium fluxes (keeping the arsenic flux constant, $5.9\times10^{-6}$ mbar) on Si (111) substrates using silver as catalysts. By increasing the indium flux (indium fluxes in **c-f** are $2.0\times10^{-7}$ mbar, $4.9\times10^{-7}$ mbar, $7.4\times10^{-7}$ mbar, $9.3\times10^{-7}$ mbar, respectively), the dimension of InAs can be tuned directly from 1-D to 2-D. For all the samples, the growth time is 40 min and the growth temperature is 505 °C.

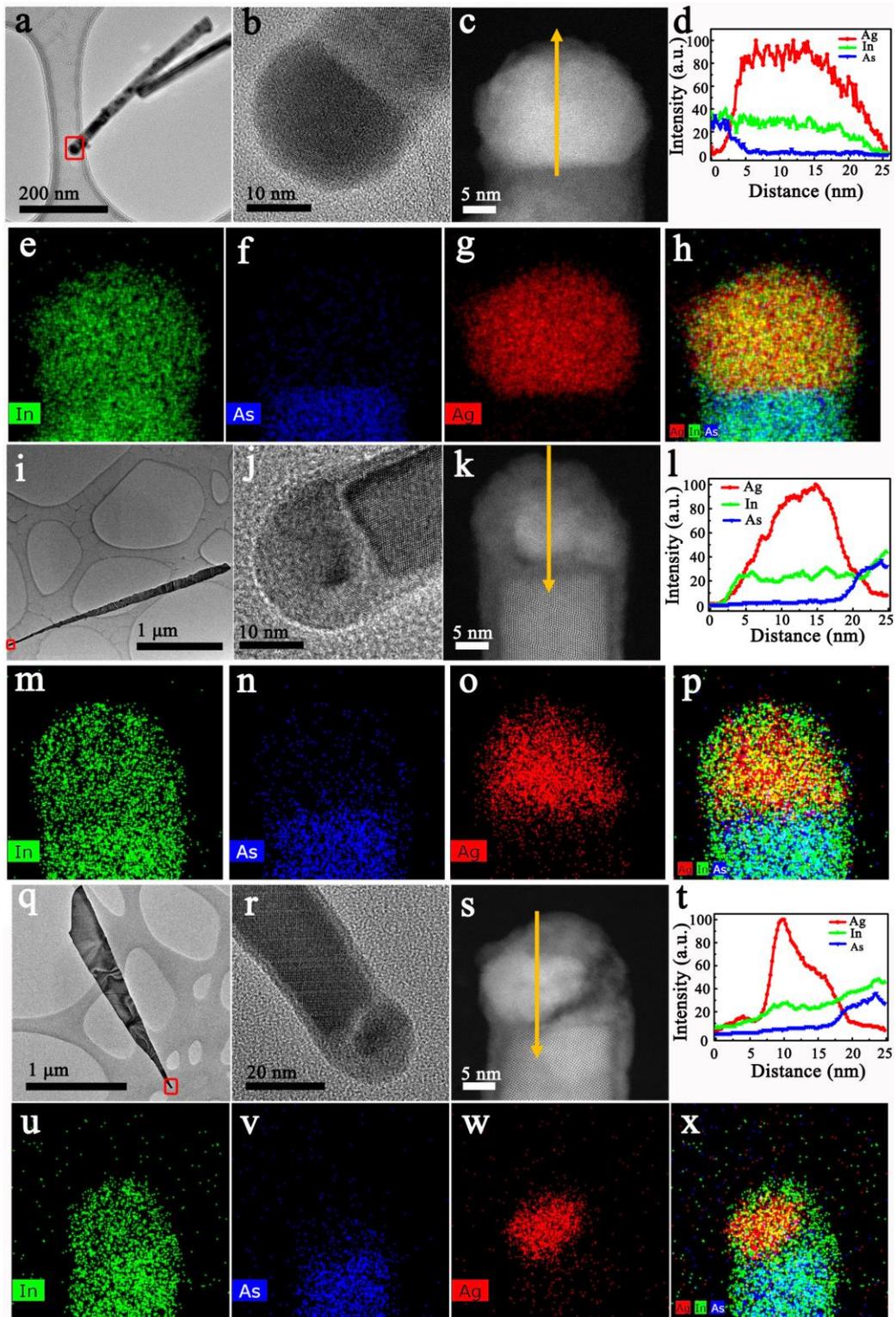

**Figure 2 | Crystal structure and elemental distribution of catalyst alloy particle of the InAs with the dimensional evolution from 1-D to 2-D**. **a-c**, **i-k**, **q-s**, TEM, high-resolution TEM and HAADF-STEM images taken from the InAs nanostructures grown with the indium flux of $2.0\times10^{-7}$ mbar, $4.9\times10^{-7}$ mbar and $7.4\times10^{-7}$ mbar, respectively. The red rectangles in **a**, **i**, **q** highlight the regions where the high-resolution TEM images were recorded. **e-h**, **m-p**, **u-x**, False-color EDS elemental maps taken at the top region of the InAs nanostructures in (c), (k) and (s), respectively. **d**, **l**, **t**, EDS line scans taken along the axial direction (marked with yellow arrows) of the catalyst alloy particle in (c), (k) and (s), respectively.

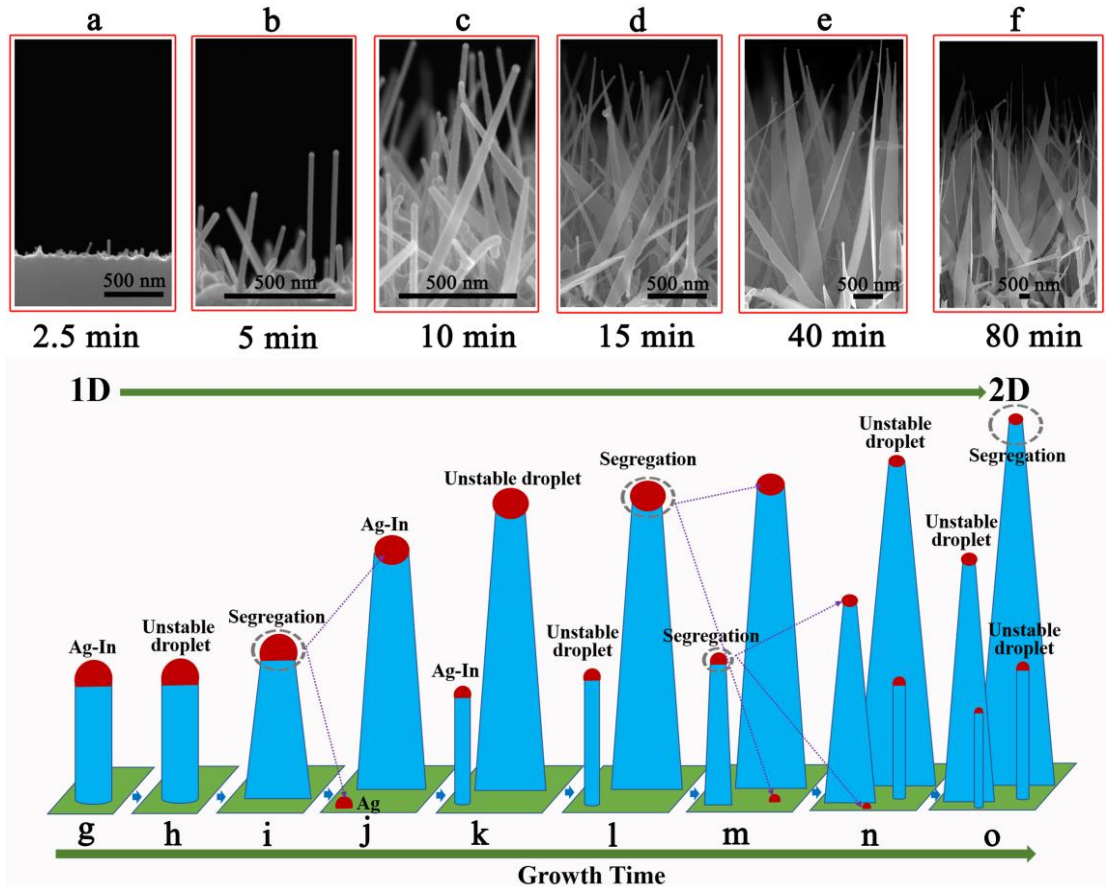

**Figure 3 | Experimental results of the InAs grown under the same indium-rich condition with different growth time and schematic demonstration of the catalyst alloy segregation process for the dimensional evolution of InAs from 1-D to 2-D. a-f**, Side-view SEM images of InAs grown on Si (111) substrates with growth time of 2.5 min, 5 min, 10 min, 15 min, 40 min and 80 min, respectively. For all the samples, the V/III BEP ratio is 6.3 (indium and arsenic fluxes are $9.3\times10^{-7}$ mbar and $5.9\times10^{-6}$ mbar, respectively) and the growth temperature is 505 °C. **g-o**, Schematic illustrating the catalyst alloy segregation process. **g**, An InAs nanowire with a spherical silver-indium alloy droplet on its top. **h**, Silver-indium alloy droplet becomes unstable under an indium-rich growth condition. **i**, Silver-indium alloy droplet starts segregation and the morphology of InAs evolves from 1-D nanowire to 2-D nanosheet gradually. **j**, Silver-indium alloy droplet finishes segregation and the segregated silver droplet in **i** migrates to the substrate surface. **k**, Silver-indium alloy droplet on the 2-D nanosheet becomes unstable, and the segregated silver droplet seeds a new InAs nanowire growth. **l**, Silver-indium alloy droplet on the 2-D nanosheet starts segregation again, and silver-indium alloy droplet on the new 1-D InAs nanowire becomes unstable. **m**, Silver-indium alloy droplet on the 2-D nanosheet finishes segregation and the segregated silver droplet in **l** migrates to the substrate surface. Silver-indium alloy droplet on the new 1-D InAs nanowire starts segregation and the morphology of the new InAs evolves from 1-D nanowire to 2-D nanosheet gradually. **n**,**o**, Silver-indium alloy droplets undergo the same segregation process in **g-m** and high density 2-D InAs nanosheets form on the substrate.

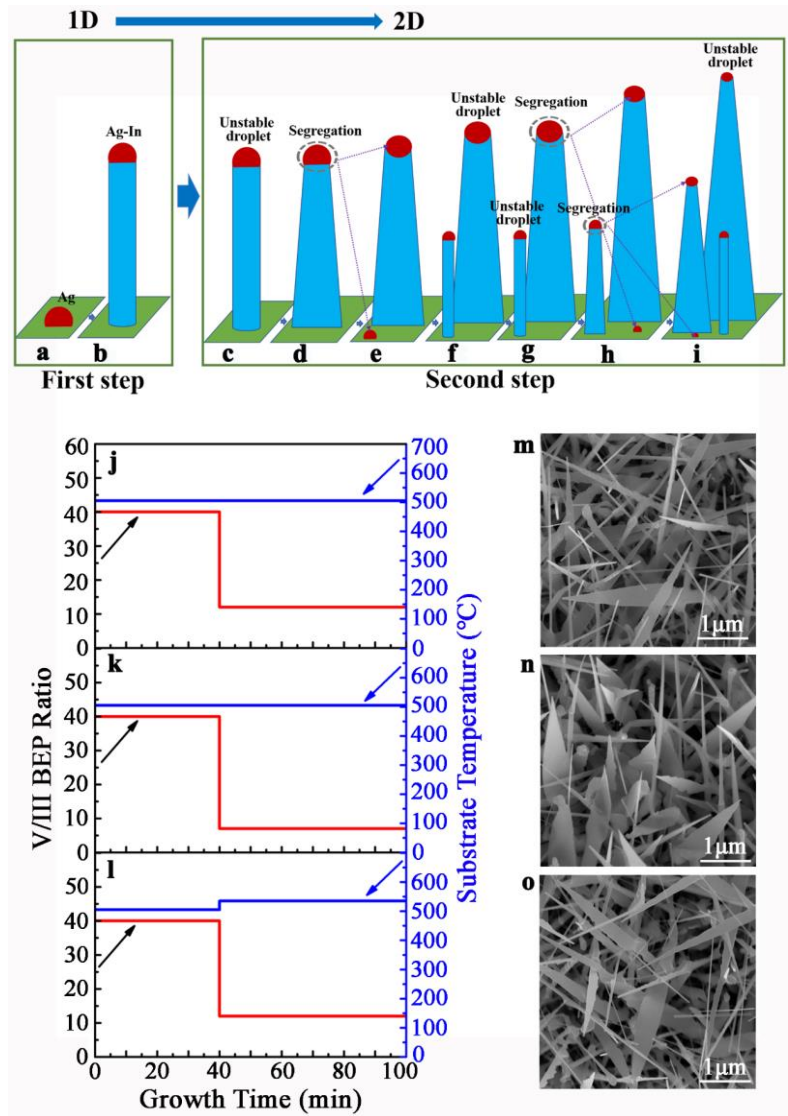

**Figure 4 | Schematic demonstration of the dimensional tunability of InAs from 1-D to 2-D using 'catalyst alloy segregation' with a two-step growth procedure and the corresponding experimental results.** First step: the 1-D InAs nanowire growth. **a**, Silver nanoparticle forms on the substrate surface. **b**, Silver nanoparticle seeds InAs nanowire growth. Second step: After the 1-D InAs nanowire growth, keeping the arsenic flux constant and increasing the indium flux, the 2-D InAs nanosheet grown under an indium-rich growth condition with 'catalyst alloy segregation'. **c**, Silver-indium alloy droplet becomes unstable under an indium-rich growth condition. **d**, Silver-indium alloy droplet starts segregation and the morphology of InAs evolves from 1-D nanowire to 2-D nanosheet gradually. **e**, Silver-indium alloy droplet finishes segregation and the segregated silver droplet in **d** migrates to the substrate surface. **f**, Silver-indium alloy droplet on the 2-D nanosheet becomes unstable and the segregated silver droplet seeds a new InAs nanowire growth. **g**, Silver-indium alloy droplet on the 2-D nanosheet starts segregation again and silver-indium alloy droplet on the new 1-D InAs nanowire becomes unstable. **h,i**, Silver-indium alloy droplets undergo the same segregation process in **c-g** and Fig. 3a-g and high density 2-D InAs nanosheets form on the substrate. **h-l**, Detailed growth conditions for the 2-D InAs nanosheet grown with the two-step growth procedure. **m-o**, 25° tilted SEM images of the 2-D InAs nanosheets grown with the two-step growth procedure. In the first step, all the InAs nanowires are grown with a V/III BEP ratio of 39.3 (indium and arsenic fluxes are $1.5 \times 10^{-7}$ mbar and $5.9 \times 10^{-6}$ mbar, respectively), growth time of 40 min and growth temperature of 505 °C. In the second step, indium fluxes are $4.9 \times 10^{-7}$ mbar (j), $8.4 \times 10^{-7}$ mbar (k), and $4.9 \times 10^{-7}$ mbar (l), and corresponding growth temperatures are 505 °C (j), 505 °C (k) and 525 °C (l); samples are all grown with 60 min.

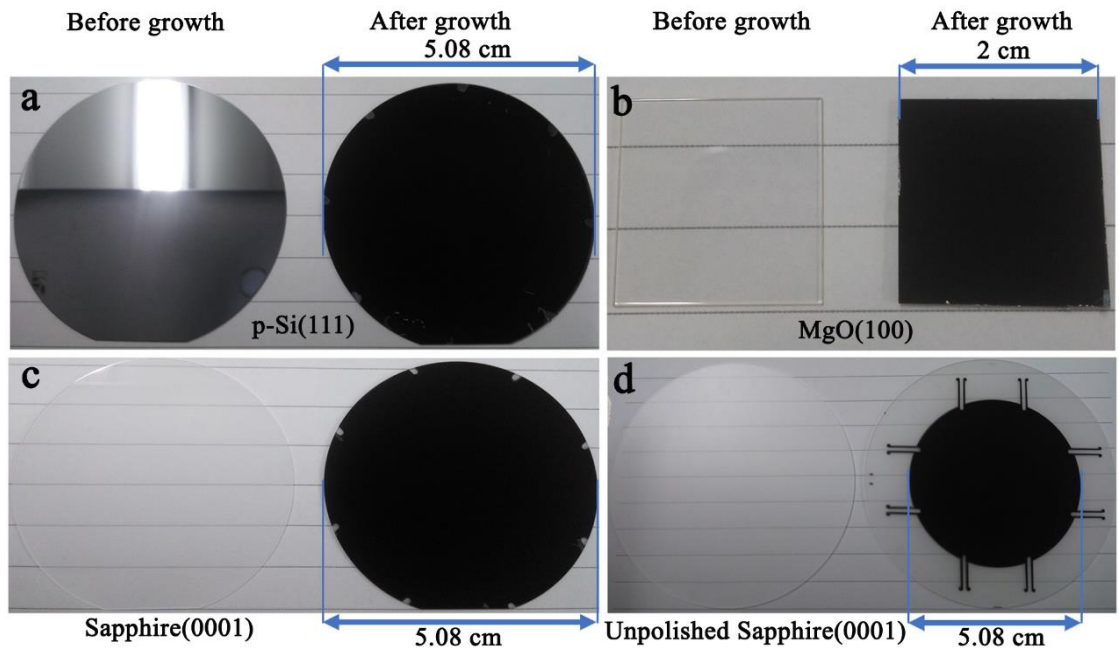

**Figure 5 | Wafer-scale 2-D InAs nanosheets grown on various substrates. a-d**, 2-D InAs nanosheets grown on 2-inch p-Si (111), 2 cm × 2 cm MgO (100), 2-inch sapphire (0001) and 3-inch (the substrate window is 2 inch) unpolished sapphire (0001) substrates, respectively. Left and right images for each panel are the substrates images taken before and after sample growth with the camera of the Redmi 5 Plus, respectively.

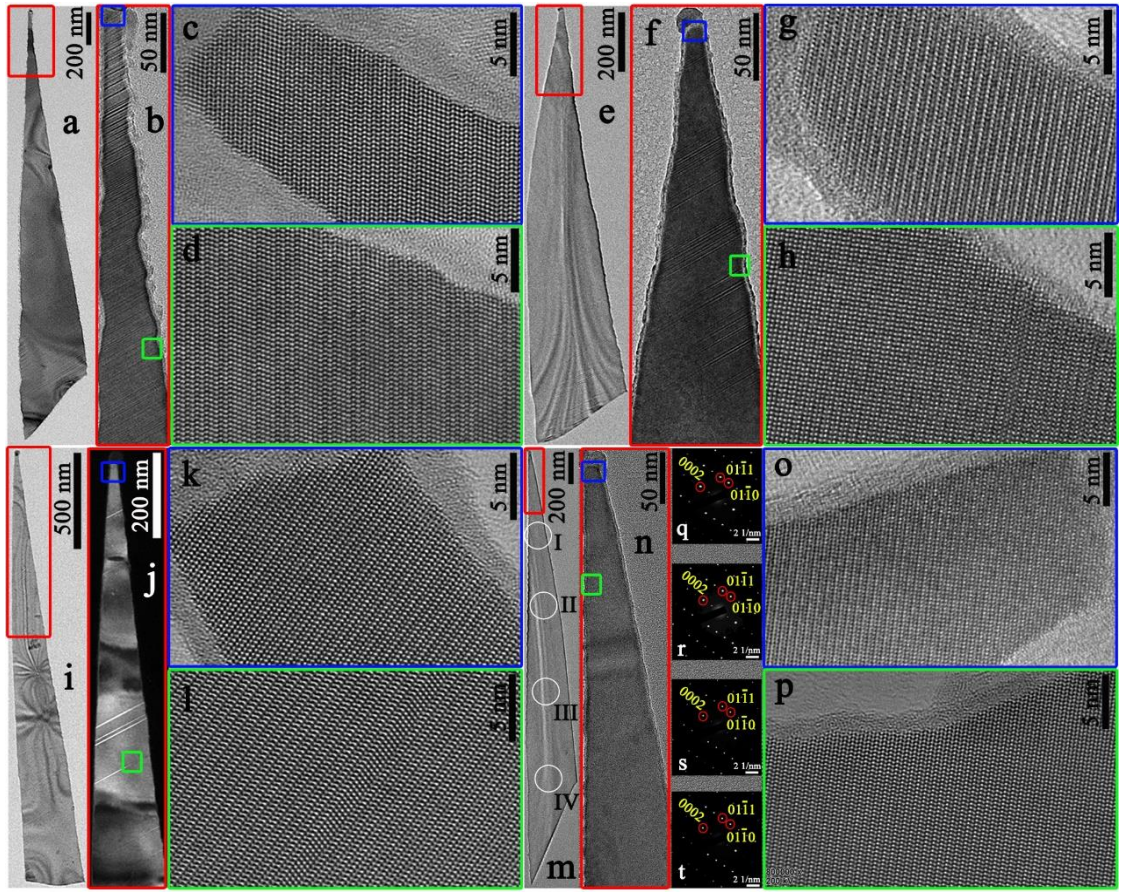

**Figure 6 | Influence of temperature on 2-D InAs nanosheet crystal phase purity. a-d**, TEM results of a typical InAs nanosheet grown at 475 °C. **a** is the overview TEM image of the nanosheet. **b-d** are high-resolution TEM images taken from different regions of the InAs nanosheet. **e-h**, TEM results of a typical InAs nanosheet grown at 505 °C. **e** is the overview TEM image of the nanosheet. **f-h** are high-resolution TEM images taken from different regions of the InAs nanosheet. **i-l**, TEM results of a typical InAs nanosheet grown at 525 °C. **i** is the overview TEM image of the nanosheet. **j**, A dark-field TEM image taken in the top section in the nanosheet. **k,l** are high-resolution TEM images taken from different regions of the InAs nanosheet. **m-t**, TEM and SAED results of a typical InAs nanosheet grown at 545 °C. **m** is the overview TEM image of the nanosheet. **n-p** are high-resolution TEM images taken from different regions of the InAs nanosheet. **q-t**, SAED patterns taken along the [2-1-10] axis from regions I, II, III, IV and V, respectively. The white circles highlight the regions where SAED patterns were recorded. All the rectangles in **a-n** highlight the regions where the high-resolution TEM images were recorded.

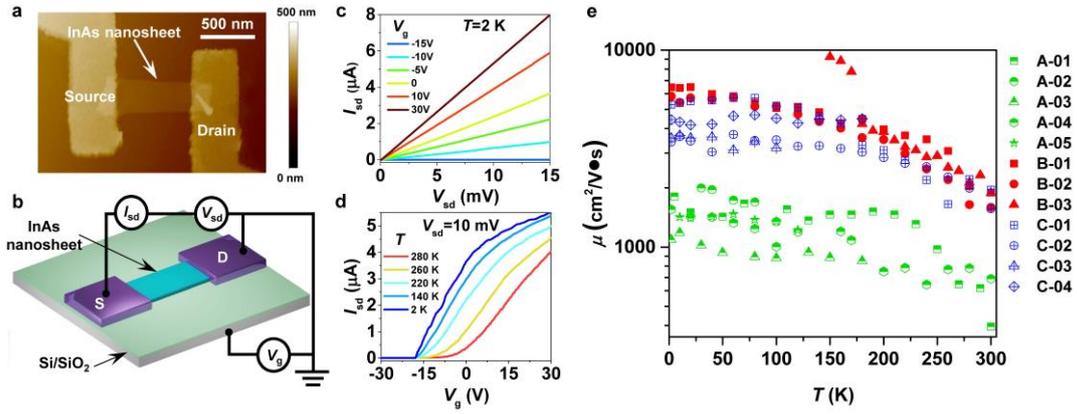

**Figure 7 | Transport characteristics of the InAs nanosheet devices. a**, AFM image of device B-01. **b**, Schematic diagram of an InAs nanosheet device. A bias voltage $V_{sd}$ is applied on the source with the drain grounded. A back-gate voltage $V_g$ is applied to tune the electron density in InAs nanosheet. **c**, Output characteristic curves of device B-01 at different back-gate voltages at $T$=2 K. **d**, Transfer characteristic curves of device B-01 as a function of back-gate voltage $V_g$ at different temperatures with $V_{sd}$= 10 mV. **e**, Extracted mobility of InAs nanosheet devices at different temperatures by fitting transfer characteristic curves. A, B and C correspond to devices made from InAs nanosheet grown in different conditions. For A devices, InAs nanosheets are grown at 455 °C with a V/III BEP ratio of 6.3. For B devices, the growth temperature is increased to 545 °C and the V/III BEP ratio is fixed at 6.3. For C devices, InAs nanosheets are grown at 505 °C with a V/III BEP ratio of 5.4.

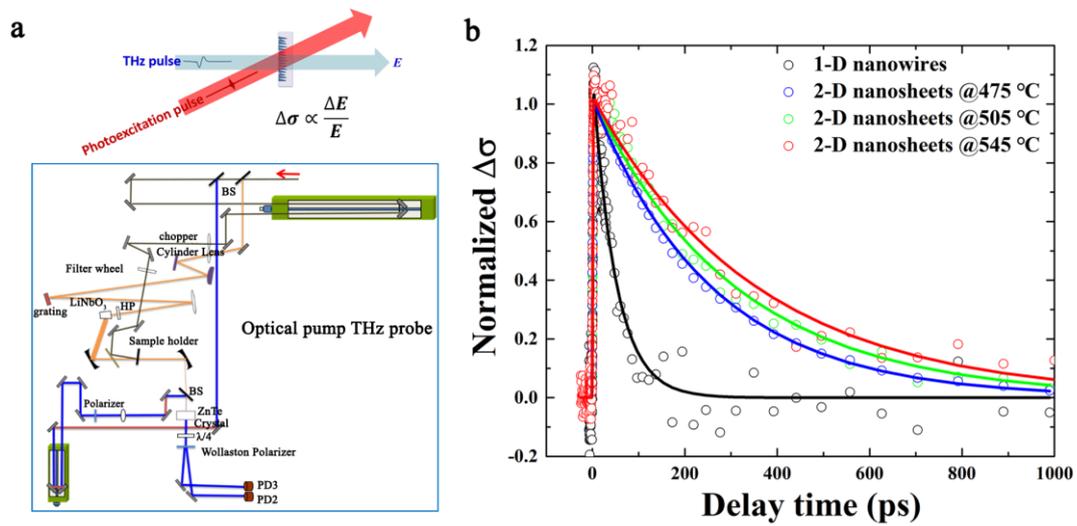

**Figure 8 | Optical properties of 1-D InAs nanowires and 2-D InAs nanosheets. a**, The setup for the time-resolved THz measurements. **b**, TRTS kinetics of 1-D InAs nanowires (black circle) and 2-D InAs nanosheets grown at 475 °C (blue circle), 505 °C (green circle) and 545 °C (red circle) after photoexcitation at 1.55 eV. The excitation photon flux is ~$2\times10^{12}$ cm$^{-2}$. The solid lines are the fitting curves by using single exponential decay functions.